\RequirePackage{lineno} 

\documentclass[preprint2]{aastex}

\shorttitle{Solar radio burst statistics}
\shortauthors{Saint-Hilaire et al.}

\begin{document}

\title{Helioseismic and Magnetic Imager observations of linear polarization from a loop prominence system}

\author{Pascal Saint-Hilaire\altaffilmark{1}, Jesper Schou\altaffilmark{2}, Juan-Carlos Mart\'inez Oliveros\altaffilmark{1}, Hugh S. Hudson\altaffilmark{1,3}, S\"am Krucker\altaffilmark{1,4},
	Hazel Bain\altaffilmark{1}, S\'ebastien Couvidat\altaffilmark{5}}

\affil{ \altaffilmark{1} Space Sciences Laboratory, University of California, Berkeley, CA, USA;\\
	\altaffilmark{2} Max-Planck-Institut f\"ur Sonnensystemforschung, Justus-von-Liebig-Weg 3, 37077 G\"ottingen, Germany; \\
	\altaffilmark{3} School of Physics and Astronomy, University of Glasgow, Glasgow, UK;\\
	\altaffilmark{4} University of Applied Sciences and Arts Northwestern Switzerland;\\
	\altaffilmark{5} W.W. Hansen Experimental Physics Laboratory, Stanford University, Stanford, CA, USA}

\email{shilaire@ssl.berkeley.edu}

\begin{abstract}
	White-light observations by the Solar Dynamics Observatory's  Helioseismic and Magnetic Imager of a loop-prominence system occurring in the aftermath of an X-class flare on 2013 May 13 near the eastern solar limb show a linearly polarized component, reaching up to $\sim$20\% at an altitude of $\sim$33 Mm, about the maximal amount expected if the emission were due solely to Thomson scattering of photospheric light by the coronal material. The mass associated with the polarized component was 8.2$\times$10$^{14}$ g.
At 15 Mm altitude, the brightest part of the loop was 3($\pm$0.5)\% linearly polarized, only about 20\% of that expected from pure Thomson scattering, indicating the presence of an additional unpolarized component at wavelengths near Fe I (617.33 nm), probably thermal emission. We estimated the free electron density of the white-light loop system to possibly be as high as 1.8$\times$10$^{12}$ cm$^{-3}$.
\end{abstract}


\section{Introduction}
	\citet{Martinez2014} (thereafter Paper I) have already reported on the observations by the Helioseismic and Magnetic Imager
	\citep[HMI,][]{Schou2012b,Scherrer2012} of the Solar Dynamics Observatory \citep[SDO,][]{Pesnell2012} of 
coronal emission from two flares occurring on 2013 May 13. 
	Both of these also showed white-light (WL) footpoint sources at the level of the photosphere. 
	The gradual coronal emissions can be identified as visual counterparts of the classical loop-prominence system, but were brighter than expected and possibly seen in the continuum rather than line emission, as inferred from high-resolution HMI spectra. 
	In this interpretation, the coronal sources detected by HMI in these flares represent flare loops, initially heated to X-ray temperatures, and detected in the process of cooling. 
	The authors found the HMI flux to exceed the radio/X-ray interpolation of the bremsstrahlung produced in the flare soft X-ray sources by at least one order of magnitude, 
	implying the participation of cooler sources that could produce free-bound continua and possibly line emission detectable by HMI.

	\begin{figure*}[ht!]
	\centering
	\includegraphics[width=16.6cm]{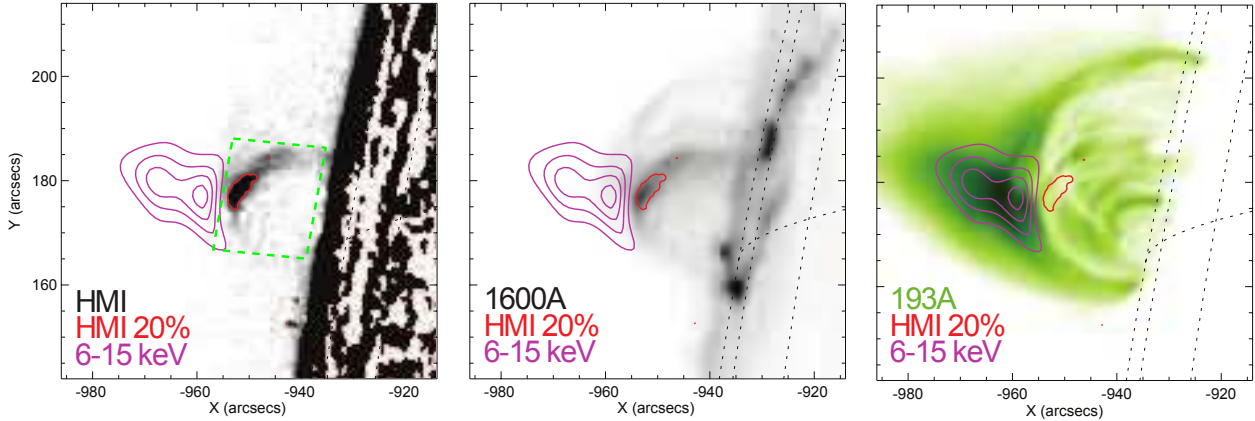}
	\caption{The gradual-phase sources, with image times: 16:25:22.7 (WL), 16:25:28.1
		(160 nm), and 16:25:21.5 (19.3 nm). The AIA 19.3 nm image shows loop-shaped absorption features,
		corresponding loosely to the WL and UV loops. 
		The purple contours are from RHESSI observations of thermal X-rays, and the red line is the 20\% contour level of the HMI image. 
		Reproduction of Fig. 5 from Paper I, with the addition of a green box used to accumulate the time profiles displayed in Figure~\ref{fig:HMIlc}.}
	\label{fig:FigX}
	\end{figure*}

Historically, the loop-prominence phenomenology played a major role in 
establishing the standard model of large-scale magnetic reconnection as a 
mechanism for the formation of flare loops projected against the corona, 
initially through observations in chromospheric lines \citep[e.g.,][]{Svestka1972}.
Observations of polarization in broad-band continuum by coronagraphs typically
do not extend low enough to study the arcade development; for example, 
the Mk4 K-coronameter at Mauna Loa only observes above about 1.12 
R$_\odot$, some 80 Mm above the photosphere.
Nevertheless some direct broad-band intensity observations in the lower
corona have been reported (Hiei et al., 1992; Leibacher et al., 2004).
These did not include the polarization signatures that HMI provides.

	In this paper, we will concentrate on one of the two events discussed in Paper I, SOL-2013-05-13T16:01 (X2.8), 
	and we present evidence of a Thomson-scattered component in the HMI emission, which allows us to discriminate between emission mechanisms contributing to the observed HMI emission. We describe the interpretation of the polarization signatures in the Appendix.
	
		\begin{figure}[ht!]
		\centering
		\includegraphics[width=7.5cm]{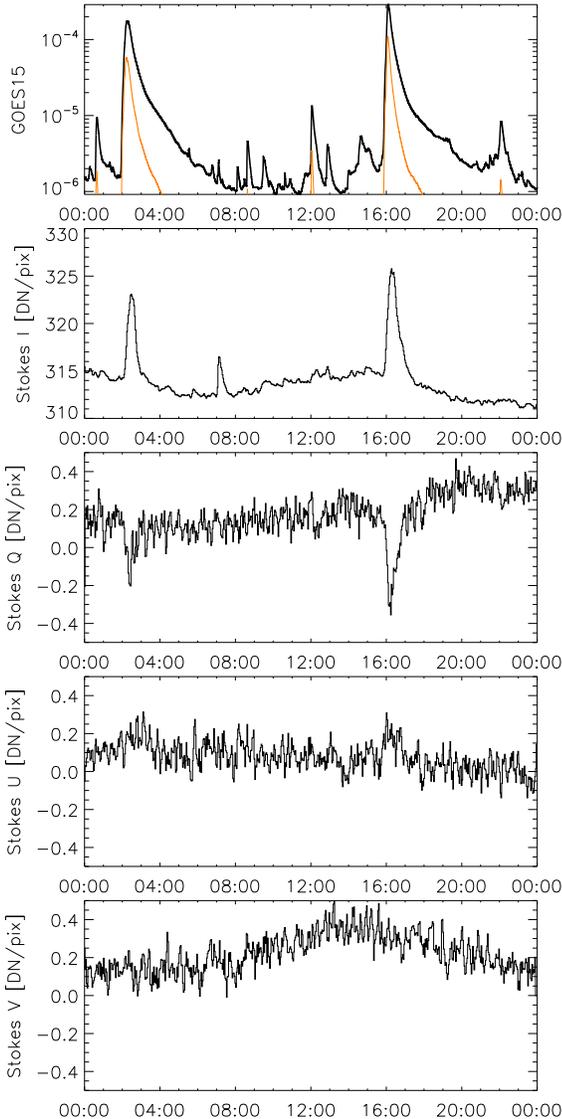}
		\caption{
			2013 May 13 GOES X-ray (black: 1--8 \AA\ ; orange: 0.5--4 \AA\ ) and HMI Stokes I, Q, U, V time profiles.
			The HMI lightcurves are generated by averaging over all six filters the pixels in the green box of Figure~\ref{fig:FigX}.
			Each pixel within the box was temporally median-filtered (with window size 3). No other data alterations were performed.
		}
		\label{fig:HMIlc}
		\end{figure}

\section{HMI observation scheme}
	HMI's side camera provides full Stokes information every 135 s in six wavelengths spanning the range $\pm$~17.25 pm around the photospheric Fe I (617.33 nm) line \citep[see][for details]{Schou2012a,Schou2012b,Scherrer2012}.
Wavelength offsets for each of the six filters (labeled I0, I1, I2, I3, I4, \& I5) are, respectively, 17.2, 10.3, 3.4, -3.4, -10.3, \& -17.2 pm, corresponding to Doppler shifts of 8.3, 5.0, 1.65, -1.65, -5.0, and -8.3 km/s (with redshifts being positive).
	For each filter, the Stokes I+Q, I-Q, I+U, I-U, I+V, I-V fluxes are the observables, with cross-talk less than the 1\% level \citep{Schou2012a}.
	These observables can easily be linearly-combined into the regular Stokes I, Q, U, and V components, at least for events which evolve relatively slowly compared to the 135 s cadence. This is the case for the loops we have observed after an X-flare on the solar east limb SOL2013-05-13T16:01 (X2.8), described in Figure~\ref{fig:FigX}.

\section{Observations} \label{sect:obs}

	Figure~\ref{fig:FigX} is a reproduction of Figure 5 of Paper I.
	It displays the flare WL loop (left), the UV loop (middle) at a similar location, and the EUV loop system in the background (right).
	The latter is clearly different from the the WL/UV loop.
	Note that we commit an abuse of language by calling the HMI observations ``white-light": the latter typically refers to broadband emission, 
	whereas HMI observes six narrow bands in the near wings of the Fe I line.
	We keep the term ``white-light", as the HMI observations described here 
definitively detect a substantial continuum contribution.
	
		Figure~\ref{fig:HMIlc}  displays 24 hour-long GOES X-ray lightcurves, as well as time profiles of the Stokes parameters averaged over all six HMI filters.
		These provide an excellent overview of the data quality and various effects that will be discussed.
		The slowly varying Stokes I behaviour can in good part be explained by the combination of varying line-of-sight (LOS) velocities of SDO with respect to the solar east limb. This, and other slowly-varying components are of negligible consequence to the analysis in this paper, which is carried out differentially.

		Of interest are the abrupt changes in Stokes I during times of strong flare activity, i.e. around 02:00 UT and 16:00 UT. 
		We will concentrate on the latter, as it was the stronger of the two.
		Notice the changes in Stokes Q and U at the same time.
		The Stokes V changes, if any, are not well observed in the presence of the noise.
	
\begin{figure*}[ht!]
		\centering
		\includegraphics[width=15.2cm]{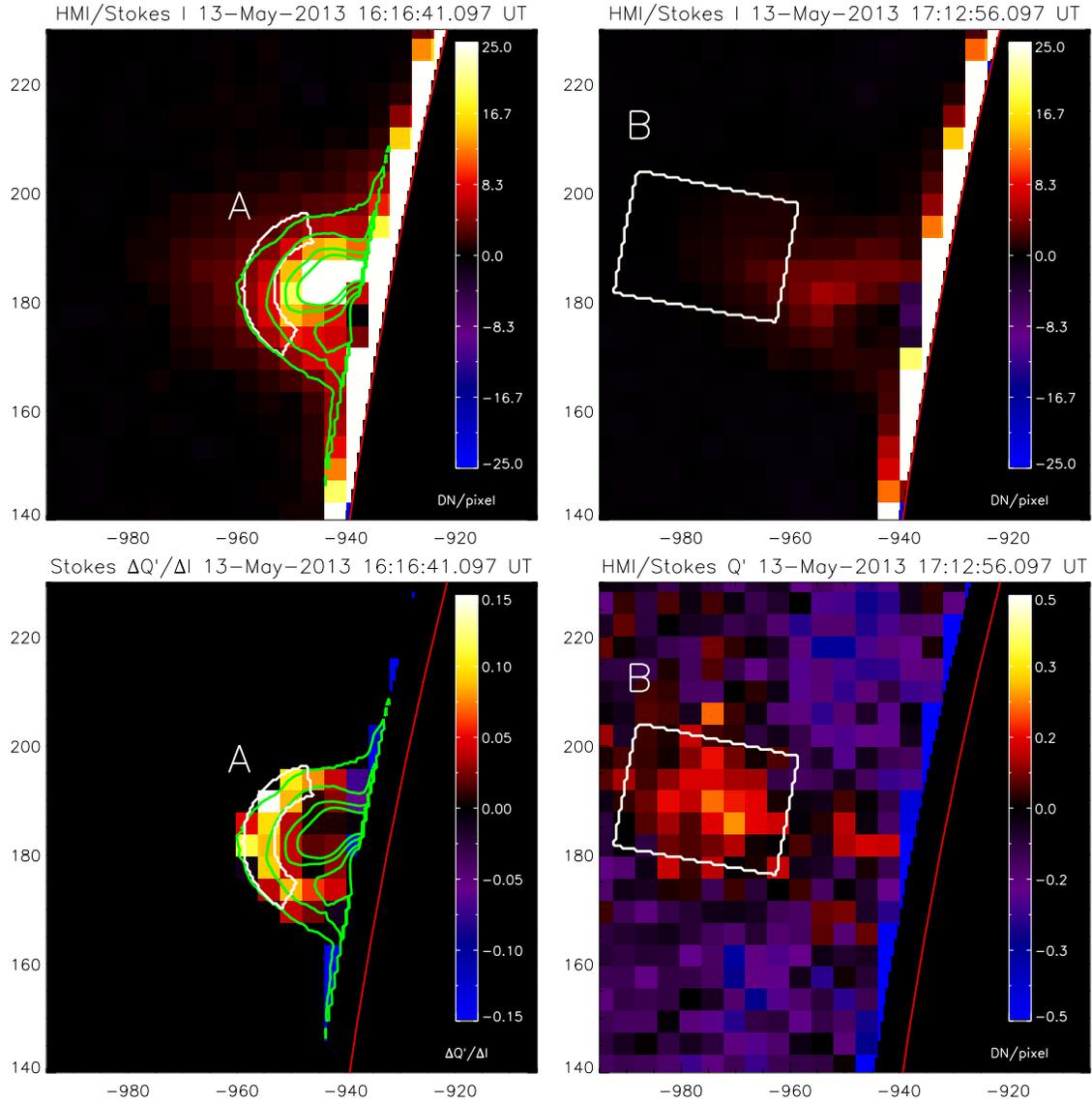}
		\caption{
			Left column: 4.5 minute average Stokes I (top left) and Stokes Q$'$/I (bottom left) centered on 16:16:41 UT.
			Right column: average from 16:03  to 18:23 UT of Stokes I (top right) and Stokes Q$'$ (bottom right) fluxes.
			The images are averages from all six HMI filters, with pre-flare images subtracted in the case of Stokes I images.
			To improve SNR, all images were rebinned to 4$''$ pixels from the original 0.5'' pixels.
			The green contours in the images on the left correspond to the 4, 8, 16, and 24 DN/pixel levels of the 0.5'' Stokes I image.
			For clarity, with pixels whose Stokes I values were below 4 DN/pixel put to zero in the Stokes Q$'$/I image,
			the areas below the photospheric limb were also displayed as black (and 5'' higher for the bottom images).
			The annular sector A and box B are the accumulation areas for the time profiles in Figure~\ref{fig:ROI_lc}.
		}
		\label{fig:IQ_img_panel}
		\end{figure*}

	To further improve the signal-to-noise ratio, we not only average the data from all six filters, but we also rotate the (Q,U) Stokes fluxes into a new coordinate system (Q$'$,U$'$) where -Q$'$ is along the solar radial (and +Q$'$ along the local horizontal).
	In this new system, and assuming the source of linear polarization to be Thomson scattering, +Q' should contain all of the linearly polarized flux, and U' is expected to be zero (as we will see later, it appears to be the case within the noise). 
	Figure~\ref{fig:IQ_img_panel} displays the Stokes I, Q$'$, and Q$'$/I images for two different time intervals, clearly showing a polarized flux component at the time and location where and when the flare loop system occurs.
	Because the I+Q and I-Q filters (and other pairs) take images 3.75 s apart, a spurious polarization signature could be created if a feature brightens rapidly.
	In our case, we estimate that this effect induces a negligible 0.025\% polarization, because of the slow time variations.
	Source motions can also produce spurious polarization signals. 
	This effect can be mitigated by ensuring that no features cross the borders of the region-of-interest where pixels are summed.
	Moreover, because the I0, I1, and I3 channels take I-Q images after I+Q images, and the I2, I4, and I5 filters the other way around, any polarization signature due to motion (or brightenings, for that matter) would appear negative in one set of channels, and positive in the other, which is not the case here.

		\begin{figure*}[ht!]
		\centering
		\includegraphics[width=16.6cm]{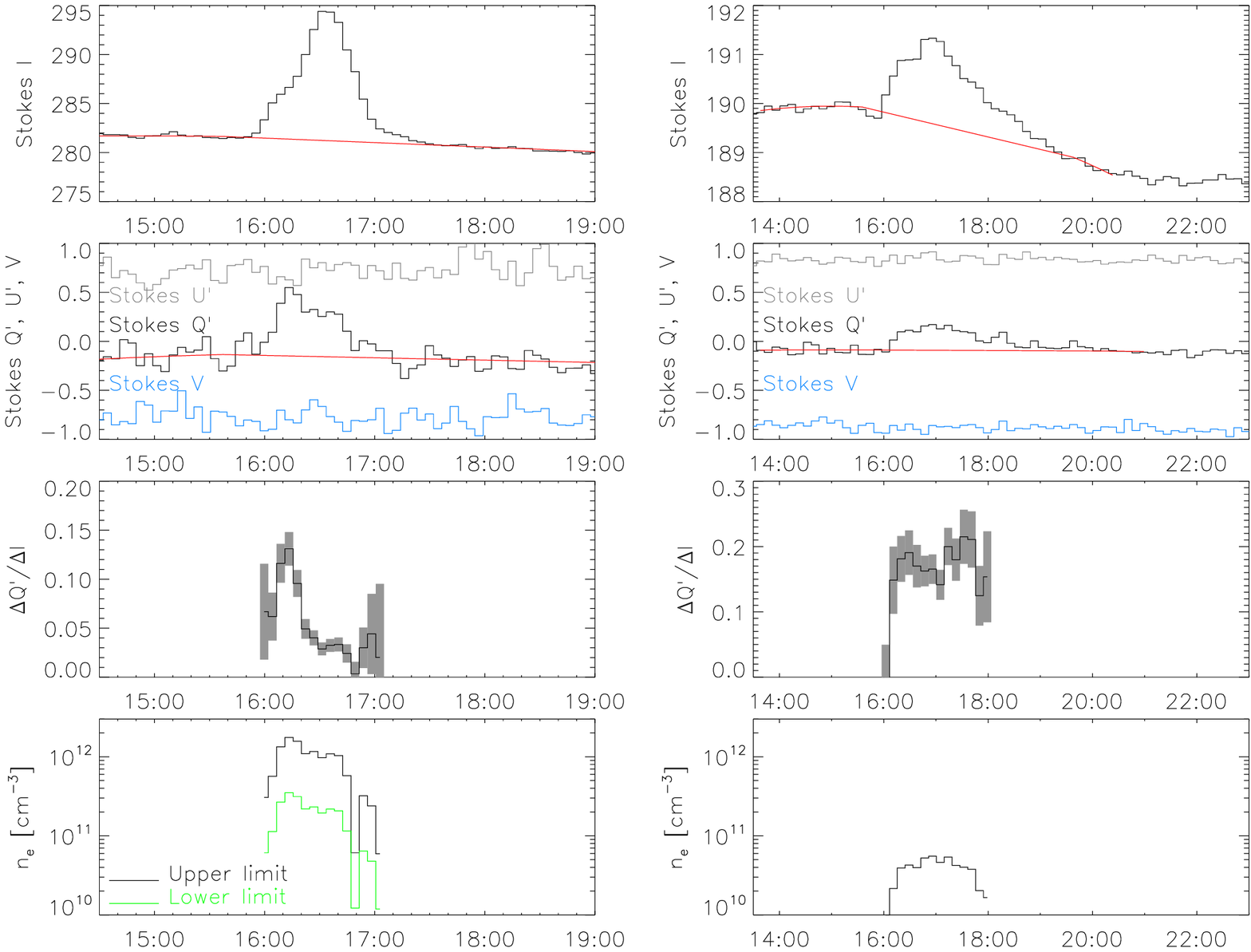}
		\caption{
			Left column: Region A (see Figure~\ref{fig:IQ_img_panel}) time profiles (4.5 mins bins).
			Right column: Region B time profiles  (9 mins bins).
			From top to bottom:
			GOES X-ray time profiles;
			Stokes I fluxes, with quadratic fit to the background (red line);
			Stokes Q', U' and V time profiles (for clarity, U' has been shifted 0.8 upwards and V 1.0 downwards);
			Stokes $\Delta$Q$'$/$\Delta$I (with 1-$\sigma$ error bars displayed in gray);
			and free electron density $n_e$.
			For clarity, the time intervals and some of the vertical scales are different from one column to the other.
		}
		\label{fig:ROI_lc}
		\end{figure*}

	In this work, we concentrate on the regions of high linear polarization fraction, and present in Figure~\ref{fig:ROI_lc} the time profiles for regions A \& B of Figure~\ref{fig:IQ_img_panel}:
	an arc-shaped area ahead of the bright loop feature around 16:17 UT (Region A), and a region of expelled material high in the corona (Region B).
	For Region A, the ratio of the change in Stokes Q$'$ and Stokes I components over background, $\Delta$Q$'$/$\Delta$I, shows a linear polarization level varying from a high of $\sim$13\% to a low of $\sim$3\%, and possibly rising again towards the end.
	The Stokes V (circular polarization) component appears consistent with zero. 
	Region B essentially displays a near-constant $\sim$20\% linear polarization fraction throughout the duration of the event.
	These results are discussed in Section~\ref{sect:discussion}.
	But before going further, and in order to better explain our observations, we provide in the next section a few details on Thomson scattering in the solar context.


\section{Thomson scattering and polarization of solar (photospheric) light \citep{Minnaert1930}} \label{sect:Minnaert}
	
	Thomson scattering is a broadband emission mechanism, with a certain level of linear polarization varying with circumstances (see below).
	The scattered intensity is proportional to the number of free electrons in the scattering element (i.e. linearly-dependent on free electron densities), and is temperature-independent.

\begin{figure}[ht!]
		\centering
		\includegraphics[width=7.5cm]{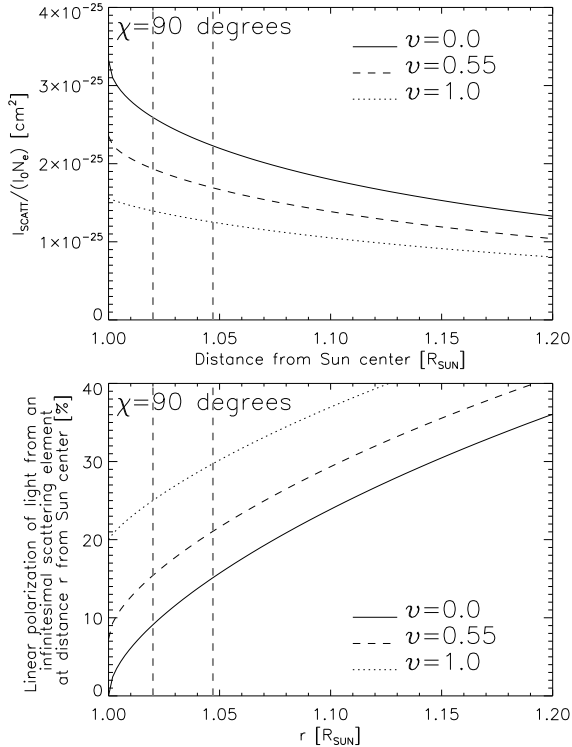}
		\caption{
			Left: Thomson-scattered intensity as  a fraction of disc-center intensity $I_0$ and free electron column density $N_e$ contained in the scattering element, for a $\chi$=90$^{\circ}$ scattering angle (as is the case at the limb).
			Right: Corresponding amount of linear polarization. The polarization angle is perpendicular to the local radial (i.e. tangential to the limb, in our case).
			$\upsilon$ is the wavelength-dependent limb darkening coefficient, with $\upsilon$=0 corresponding to no limb darkening ($\sim$IR wavelength), $\upsilon$$\approx$0.55 being  appropriate for the continuum around 617.33 nm. For completeness, we have added  $\upsilon$=0.8, corresponding to the solar 430 nm wavelength. The vertical dashed lines indicates the mean altitudes of Region A and Region B in Figure~\ref{fig:IQ_img_panel}.
		}
		\label{fig:Thomson}
		\end{figure}

	\citet{Minnaert1930} thoroughly describes the intensity and degree of linear polarization of solar light expected to be Thomson-scattered towards an observer by a scattering element in the corona, including the effects of limb darkening. Here, we have generated  plots relevant to the understanding of our specific observations: Figure~\ref{fig:Thomson} shows the relative amount of scattered flux per unit column density and fraction of linear polarization expected from an event at the limb observed at the Earth (scattering angle $\chi$$\approx$90$^{\circ}$), and for different (and wavelength-dependent) limb darkening coefficients $\upsilon$ \citep[see][for details]{Minnaert1930}. 
$\upsilon$=0 corresponds to no limb darkening, whereas $\upsilon$$\approx$0.55 approximates well the continuum near 617.33 nm.
(Because of Doppler shift we expect the wavelength of the light scattered towards HMI to have been originally far from the Fe I line.)

The light scattered towards HMI by free electrons at an altitude of 19$''$' above the limb is hence expected to possess 15.4\% linear polarization, if Thomson scattering is the only emission mechanism at work, and for $\chi$=90$^{\circ}$ scattering angle. This figure drops to 15.3\% for $\chi$=85$^{\circ}$, and 14.9\% for $\chi$=80$^{\circ}$.

\section{Discussion} \label{sect:discussion}

	At the mean altitude of Region A, a linear polarization fraction $\Delta$Q$'$/$\Delta$I$\approx$15\% is expected if the emission is entirely due to Thomson scattering.
	From the actual value of $\Delta$Q$'$/$\Delta$I in the third plot of Figure~\ref{fig:ROI_lc}, it can be deduced that 88($\pm$13)\% (around 16:10 UT), down to $\sim$20($\pm$4)\% (around 16:35 UT) of the emission is due to Thomson scattering.
	Thus, the data suggests that most or all of the emission at the forefront of the flare loop development is due to Thomson scattering,
	and there is an additional source of unpolarized emission present within the brightest feature of the loop system.
	This unpolarized component is discussed later on. For now, we concentrate on interpretating the polarized component. 

	Section~\ref{sect:Minnaert} details how the free column density can be deduced from the above results.
	The free electron column density in Region A peaks around 16:13 UT at $N_e \approx$ 3.9($\pm$0.5)$\times$10$^{20}$ cm$^{-2}$. 
	This leads in turn to a number of free electrons of 3.1($\pm$0.4)$\times$10$^{38}$, and a total mass of 5.2($\pm$0.7)$\times$10$^{14}$  g, or about a third of that of an average CME \citep{WebbHoward2012}.
	To derive free electron densities, the source volume must be inferred; for a uniform source a plausible estimate for the line-of-sight (LOS) depth $L$ comes from the square root of the area of the Stokes Q$'$ region (not shown here),
	which leads to $L$=11($\pm$2) Mm and a probable lower limit for the free electron density of $n_e$=3.5$\times$10$^{11}$ cm$^{-3}$.
	However, assuming a homogeneous source, another approximation for $L$ is suggested by the  geometry of the rising WL loop, and given by the apparent FWHM of the loop feature (Figure~\ref{fig:FigX}), i.e. $L$$\approx$2.2 Mm. 
	This in turn leads to a free electron density estimate for the loop of $n_e$=1.8($\pm$0.2)$\times$10$^{12}$ cm$^{-3}$, probably a rough upper limit for unit filling factor.
	Doing a similar analysis for the whole area beneath Region A, down to 8$''$ above the photosphere (an ad hoc value) leads to similar peak densities occurring (unsurprisingly) earlier in the flare ($\sim$16:06 to $\sim$16:11 UT).
	Such density estimates have been reported before \citep{Svestka1972,Hiei1992,CaspiLin2010}, via other methods.
	Moreover, these upper density estimates are of the same order as the density derived from thermal hard X-rays around 16:07 UT. 
RHESSI is sensitive to free electrons belonging to $\gtrsim$8 MK plasmas, while these Thomson-scattered sources observed by HMI are sensitive to all free electrons present.
A thorough investigation of the spatio-temporal correspondance between HMI- and RHESSI-observed free electrons is the topic of the next paper in this series.

	As can be seen in Figure~\ref{fig:FigX}, the WL/UV loop tends to transform into an obscuring feature in EUV.
	Various authors have been studying EUV-absorbing features to determine the properties of the absorbing plasma, particularly in the context of prominences 
	\citep[e.g.][and references therein]{Kucera1998,Gilbert2011,LandiReale2013}.
	For most of SDO's Atmospheric Imaging Assembly \citep[AIA,][]{Lemen2012} filters, EUV photoionization opacities only kick in at temperatures below $\sim$100 kK.
	In the case of AIA's 30.4 and 33.5 nm filters, it is below $\sim$40 kK. Hence, the WL/UV loop likely contains cold plasma.	
	As previously mentioned, a varying fraction of the WL emission can be explained by Thomson scattering. 
	The rest of the emission is probably due to thermal (free-free and free-bound) emission from a low-temperature (as low as a few tens of kK) plasma:
	a column emission measure of $n_e^2 L$ (where $n_e$$\approx$10$^{12}$ cm$^{-3}$ and $L$=2.2$\times$10$^8$ cm) is indeed consistent with the observed intensities.
	Such a low-temperature plasma component would not emit in X-rays, and the optically-thick microwave emission would be essentially invisible to most microwave spectrometers or interferometers, and thusly be able to explain the observations shown in Figure 3 (left) of Paper I, where microwave- and X-ray derived hot (MK) thermal fluxes seem to match, while the WL flux is an order of magnitude above expectations. Had high brightness sensitivity and high spatial resolution radio observations been available in the gradual phase of this event, and assuming negligible non-thermal emissions, it is possible that we would have been able to observe a cool optically-thick loop in front of hot flare loops.
	

	Finally, Region B displays a linear polarization fraction close to $\sim$20\% most of the time, near the 21\% (averaged over Region B) expected from pure Thomson scattering. We estimate the rising source's peak mass at 8.2$\times$10$^{14}$ g.
The density profile in Figure~\ref{fig:ROI_lc} is achieved by using the square root of the Stokes Q$'$ feature in Figure~\ref{fig:IQ_img_panel} (bottom right) for the LOS depth, leading to $L$=23($\pm$3) Mm, and a free electron density peaking at 5.9$\times$10$^{10}$ cm$^{-3}$. 
Contrary to Region A, there is no unambiguously observed signature of an emission mechanism besides Thomson scattering.

\section{Conclusions} \label{sect:ccl}
	We have reported what we believe to be the first detection of  linearly polarized scattered white-light of an evolving flare loop system in the vicinity of the Fe I line,
	and probably the first unambiguous mass estimate of such a system, owing to the linear dependence on density and temperature-independence of Thomson scattering.
	It is only due to SDO/HMI's remarkable dynamic range and polarimetric capabilities that such a faint signature of Thomson scattering could be identified. 
	This has enabled us to dynamically 1) identify the fraction of WL emission in a flare loop at the limb that is due to Thomson scattering, 
	and 2) estimate the free electron content and the mass of the scattering sources, as well as likely limits to their free electron densities.
	Such an approach brings a powerful new diagnostic tool to the study of limb flares, and possibly the flare/CME connection as well.	



\acknowledgments
PSH, JCMO, HB, SK, and HSH were supported by NASA Contract No. NAS 5-98033.
The data used here are courtesy of NASA/RHESSI, NASA/SDO and the HMI and AIA science teams.

\bibliographystyle{apj}


\end{document}